\documentclass[prc,floatfix,superscriptaddress,letter]{revtex4}

\usepackage{color}
\usepackage[normalem]{ulem}
\input epsf

\usepackage{amsmath}
\usepackage{graphicx}
\usepackage{url}
\usepackage{color}
\usepackage{subfigure}
\usepackage{longtable}
\usepackage{dcolumn}

\begin{document}

\begin{flushright}
JLAB Note JLAB-PHY-15-2132\\31-Aug-2015
\end{flushright}

\title{Search for Hidden-Charm Pentaquark with CLAS12}

\newcommand*{\JLAB}{Thomas Jefferson National Accelerator Facility, Newport News, Virginia 23606, USA}
\newcommand*{\WILL}{William I. Fine Theoretical Physics Institute, University of
Minnesota, Minneapolis, MN 55455, USA}
\newcommand*{\SCHOOL}{School of Physics and Astronomy, University of Minnesota, Minneapolis, MN 55455, USA}
\newcommand*{\ITEP}{Institute of Theoretical and Experimental Physics, Moscow, 117218, Russia}

%\newcommand*{\JLABindex}{32}
%\affiliation{\JLAB}
 %%%%%%%%%%%%%%% END OF Latex Macros for institute addresses  %%%%%%%%%%%%%%%%%%%%%%%%% 
\author {V.~Kubarovsky} 
\affiliation{\JLAB}
\author {M.B.~Voloshin} 
\affiliation{\WILL}
\affiliation{\SCHOOL}
\affiliation{\ITEP}
%\date{\today}
\maketitle

\section{Introduction}

\begin{figure}[h]
\begin{center}
\includegraphics[width=0.99\textwidth]{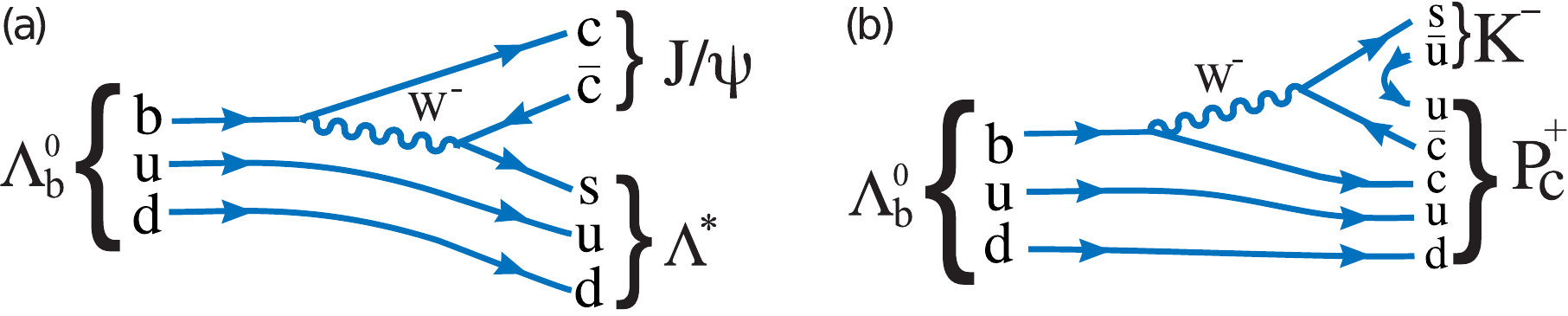}
\end{center}
\vskip -0.3cm
\caption{Feynman diagrams for (a) $\Lambda_b^0\to J/\psi \Lambda^*$ and (b) $\Lambda_b^0\to P_c^+ K^-$ decay.}
\label{Feynman-Pc}
\end{figure}

\begin{figure}[h]
\vskip -0.7cm
\begin{center}
\includegraphics[width=0.42\textwidth]{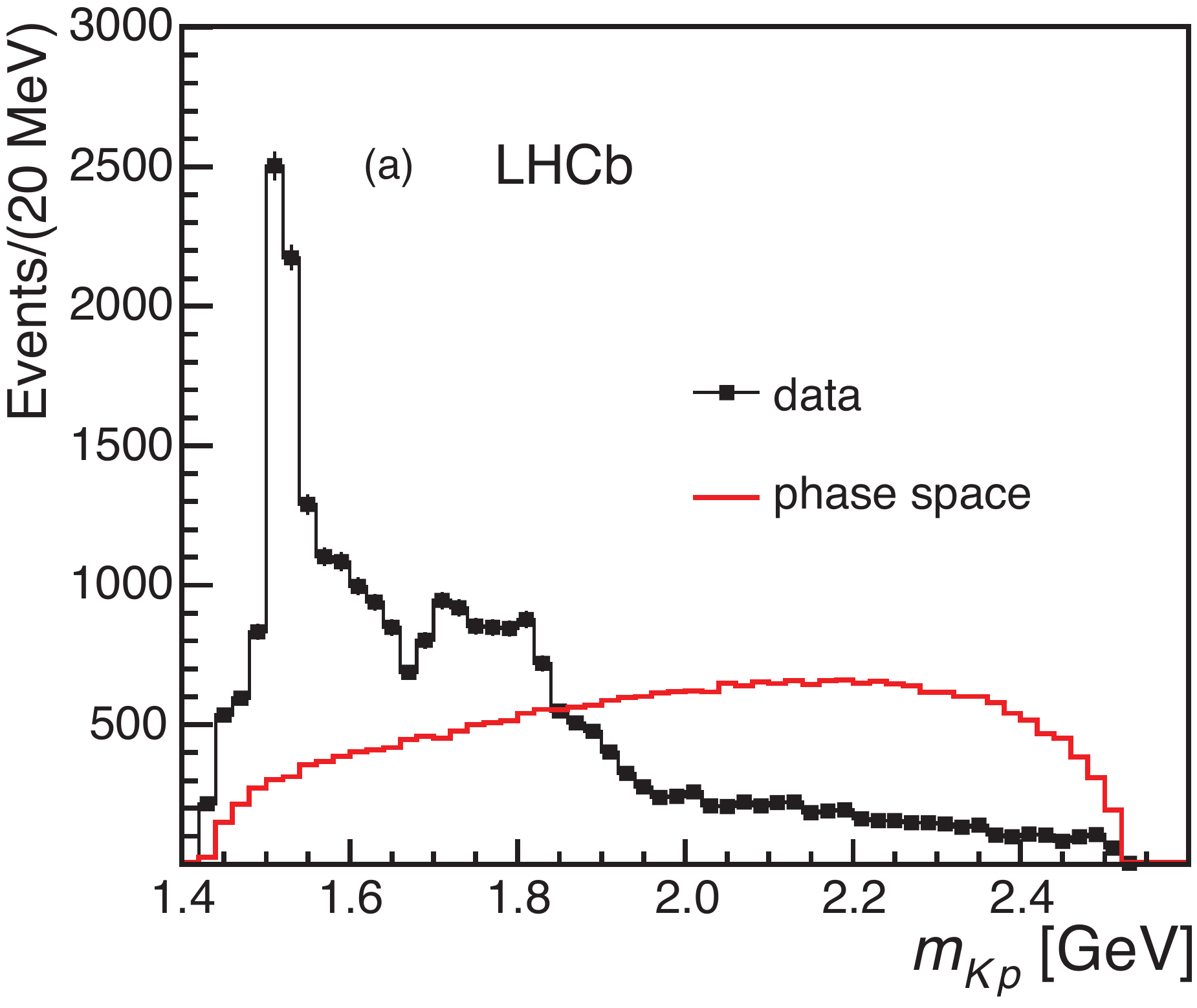}\hspace*{0.99cm}\includegraphics[width=0.42\textwidth]{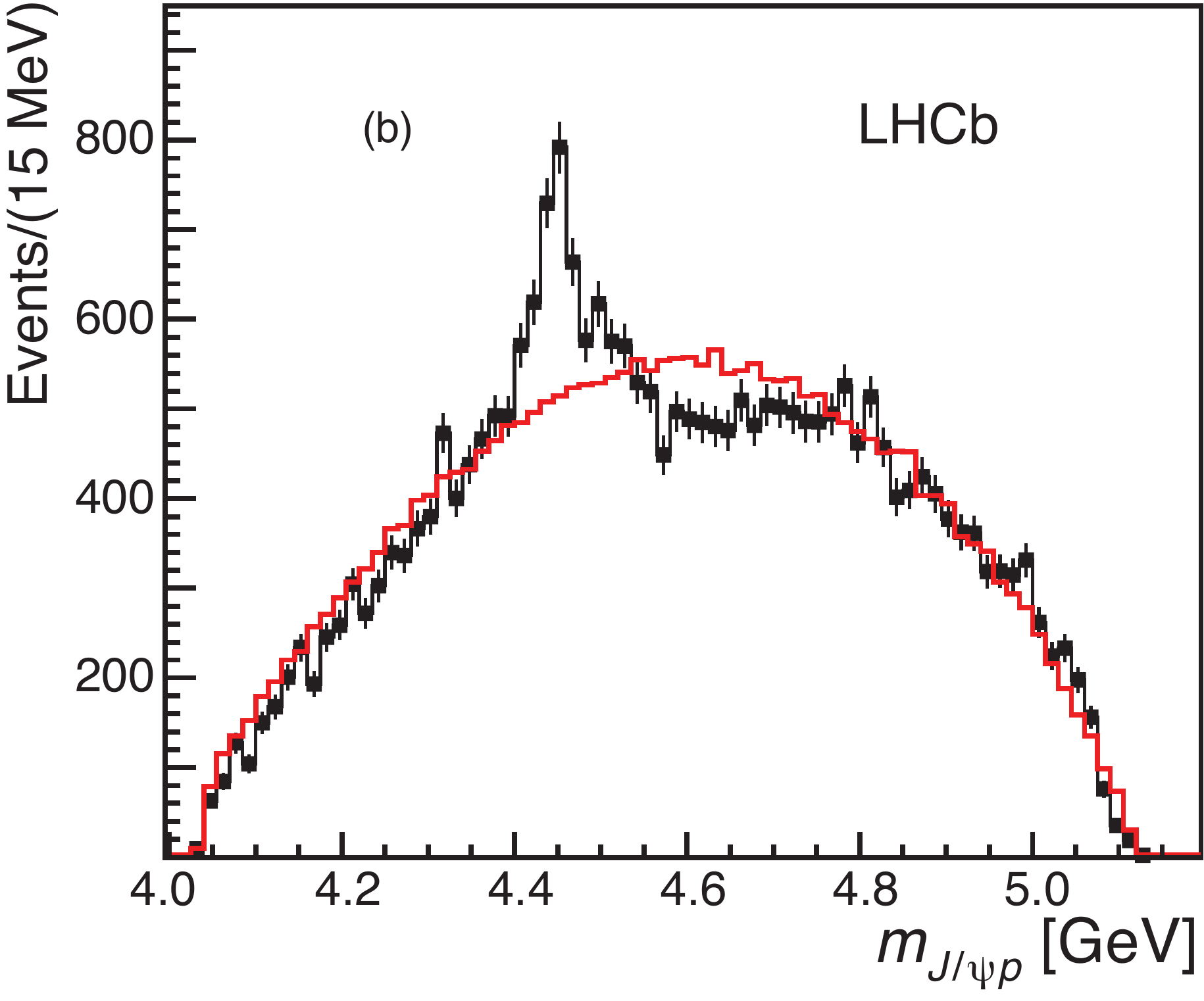}
\end{center}
\vskip -0.2cm
\caption{ Invariant mass of (a) $K^-p$  and (b) $J\psi p$ combinations from $\Lambda_b^0\to J/\psi K^-p$ decays. The solid (red) curve is the expectation from phase space.}
\label{mpk-mjpsi}
\end{figure}

LHCb recently announced the 
discovery of two exotic structures in the $J/\psi + p$ decay channel, which have been referred to as charmonium-pentaquark states~\cite{lhcb}.
They labeled these states as $P_c(4380)$ and $P_c(4450)$ and claimed that the minimum quark content is $c\bar c u u d$. 
The pentaquarks were observed in the decay $\Lambda_b^0\to K^-P_c^+$, $P_c^+\to J/\psi p$.
The Feynman diagram for the signal and background are presented in  Fig.~\ref{Feynman-Pc}.
One state has a mass of $4380\pm8\pm$29 MeV and a width of $205\pm18\pm86$ MeV, while the second is narrower, with a mass of $4449.8\pm1.7\pm2.5$ MeV and a width of $39\pm5\pm19$ MeV.  The preferred $J^P$ assignments are of opposite parity, with one state having spin $3/2$ and the other $5/2$. 
The narrow $P_c(4450)$ state is clearly visible in the $J/\psi p$ invariant mass spectrum in Fig.~\ref{mpk-mjpsi}b.
An evidence for the other state $P_c(4380)$ was found after rather complicated data analysis.

%\clearpage

The decays of conventional baryons to $J/\psi+X$ are strongly suppressed by the Okubo-Zweig-Iizuka rule. It provides a hint  that  these resonances contain a $c\bar c$ pair  and  3 light quarks in the initial state to conserve the  baryonic number.
In addition, the masses of these states ($\approx$~4.4 GeV) are close to the sum of the mass $J/\psi$ and proton. The narrow width (especially for the $P_c(4450)$) supports the hypothesis that these heavy baryonic states   have small probability to decay to the low mass mesons and baryon,
which would be very 
difficult to explain if these states consist of the light quarks only. So the interpretation of these structures as pentaquark with hidden charm looks very reasonable. 

There are  several models on the market that attempt to describe the internal structure of the pentaquarks with hidden charm.
\begin{itemize}
\item  The $P_c$ states are interpreted as a composite of the charmonium state and the proton or exited nucleon states~\cite{vpk} similar to the known resonances N(1440) or N(1520). In this case we expect that the branching ratio $Br(P_c\to J/\psi+p)$ is sizable and may lay in the range from 1\% to 10\%. It was pointed out in~\cite{vpk}  that in this model one can expect that the decays of the pentaquarks $P_c\to J/\psi+p+\pi$ and $P_c\to J/\psi+p+\pi+\pi$ should at least compete in the total rate with the observed $J/\psi+p$ channel.
\item  One of the pentaquarks, $P_c$(4450) is interpreted \cite{mo} as a $\chi_{c1} p$ 
baryo-charmonium composite resonance.
\item The $P_c$ states are hadronic molecules~\cite{cllz,cclsz,rno,he}. These molecules consist of  a charmed baryon and charmed meson which are weekly  coupled with each other.  Such pentaquarks will decay predominantly to the charmed baryon and charmed meson.
\item Pentaquarks made of tightly correlated diquarks~\cite{mpr,amnss}
or colored baryon-like and meson-like constituents~\cite{mm,lebed}.
\item It has been also suggested that at least one of the peaks is not a resonance at all, but rather a kinematical singularity due to rescattering\cite{gmwy,lwz,mikhasenko,amnss}  in the decay $\Lambda_b\to J/\psi p K^-$.
\end{itemize} 

%Clearly, resolving between the models and clarifying the nature of the hidden-charm pentaquark states, requires further experimental studies. These states were observed in the the decay mode $(J/\psi+p)$. It is natural to expect that these states can be produced in photopropduction process $\gamma p\to P_c\to J/\psi+p$ where these states will appear as s-channel resonances. The observation of the $P_c$ states as s-channel resonances will provide the strong support that these resonances are not the artifact of so called anomalous triangle singularity that create the threshold enhancement. 
%The search of the other decay modes of the $P_c$ states will provide the tools to resolve between the models and clarifying the nature of the discovered hidden-charm pentaquark. 

\begin{figure}[ht]
\begin{center}
\includegraphics[width=0.45\textwidth]{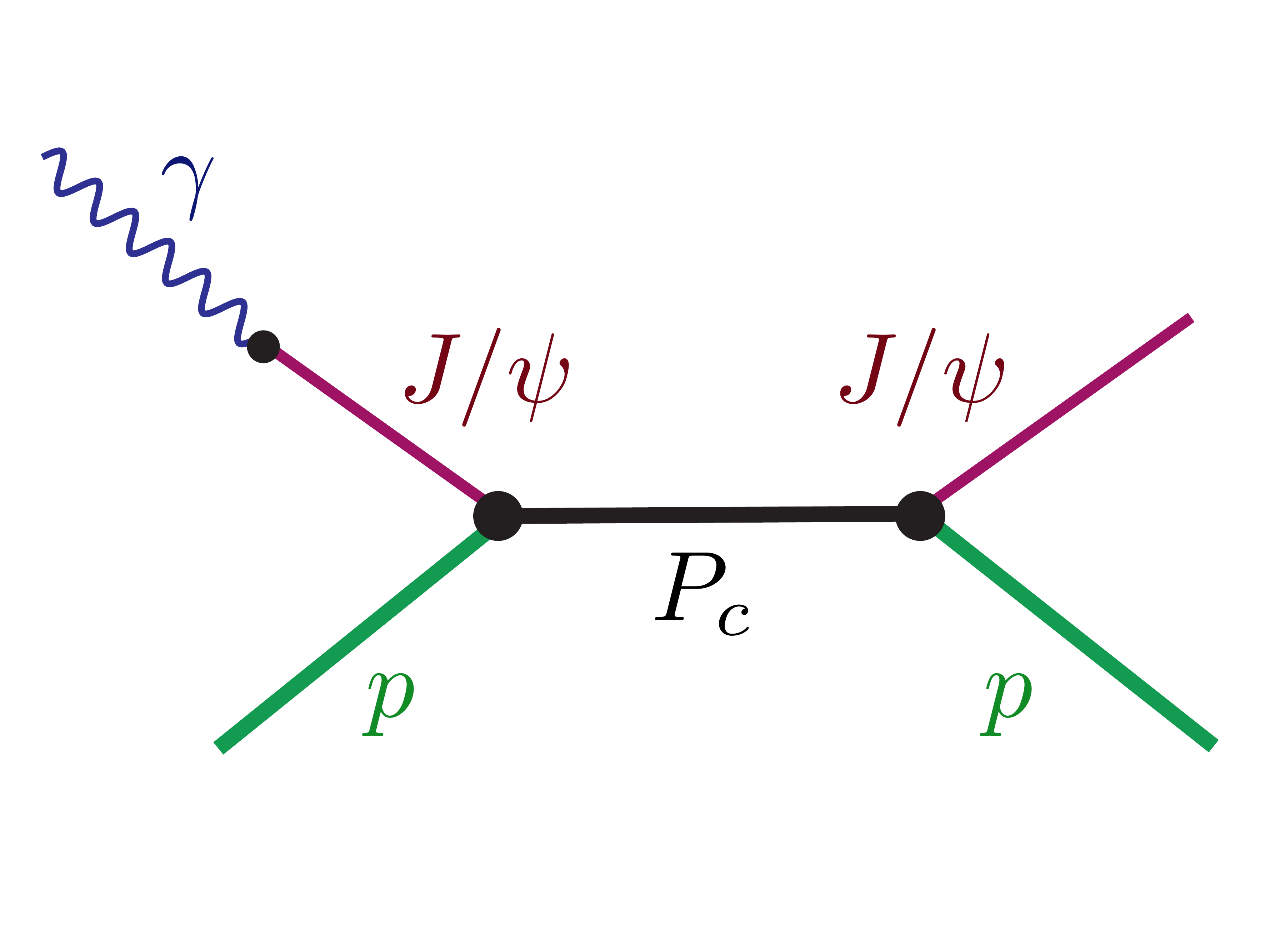}
\caption{Feynman diagram for the pentaquark photoproduction process.
}    
\label{fndiag}
\end{center}
\end{figure}

Clearly, resolving between the models and clarifying the nature of the discovered hidden-charm pentaquark peaks, and possibly searching for similar peaks with other quantum numbers, requires further experimental studies. 
These states were observed in the  decay mode $J/\psi+p$. For this reason it is natural to expect that these states can be produced in photoproduction process $\gamma+p\to P_c\to J/\psi+p$ where these states will appear as s-channel resonances at photon energy around 10\,GeV~\cite{wlz,vpk,karl}.
Fig.~\ref{fndiag} illustrates the S-channel pentaquark photoproduction process.

Such experiments can be advantageous for detailed studies of the production and decay properties of the pentaquark resonances in comparison with the LHCb environment.  The discussed yield is determined by the branching fraction $Br(P_c \to \gamma + p)$, and it was shown~\cite{vpk}  that this parameter can be expressed in terms of $Br(P_c \to J/\psi+p)$ by a relation similar to  vector dominance for the $J/\psi$ photoproduction. Although such dominance cannot be justified as a general rule, in the situation at hand it can be applied due to arguments based on the heavy quark properties and special kinematics of the processes involved. As a result the peak cross section  $\sigma(\gamma + p \to P_c \to J/\psi + p)$ is proportional to $Br^2(P_c \to J/\psi + p) $ and  can reach tens of nanobarns or more, if $Br(P_c \to J/\psi + p) \sim 1-10\%$. Such relatively large cross section may allow fairly detailed studies of the pentaquarks and a search for other similar states. In particular, it may be realistic to study the decays of the $P_c$ states into 
$J/\psi~ p \pi$ and $J/\psi~ p \pi \pi$. These types of  decays should be prominent, if the $P_c$ states are dominantly a baryo-charmonium, i.e. a hadro-quarkonium, type~\cite{mv07,dv} composite of $J/\psi$ and excited nucleon states similar to the known resonances $N(1440)$ and $N(1520)$. Such pattern of the decays of the $P_c$ resonances would disfavor the molecular models~\cite{cllz,cclsz,rno,he}, where one would expect the natural decay channels into a charmed hyperon and a meson, or from the $\chi_{c1} p$ complex model~\cite{mo}, where the expected dominant decay is $P_c(4450) \to \chi_{c1}+p$. Naturally, any observation of the $P_c$ peaks in the $\gamma p$ cross section would strongly disfavor the interpretation~\cite{gmwy,lwz,mikhasenko,amnss} in terms of `accidental' singularities in the $\Lambda_b$ decays.

\section{Photoproduction cross section}

For a resonance $P_c$ in the $s$ channel the photoproduction cross section of the reaction 
$\sigma(\gamma + p \to P_c \to J/\psi + p)$
is given by the standard Breit-Wigner expression (see e.g. in Ref.~\cite{pdg}, Sec.~48.1.)
\begin{equation}
\sigma(W) = { 2 J +1 \over 4} \, {4 \pi \over k^2} \, {\Gamma^2/4 \over (W-M_c)^2 + \Gamma^2/4} \, Br(P_c \to \gamma + p) \, Br(P_c \to J/\psi +p)~,
\label{bw}
\end{equation}
where $J$ is the spin of the $P_c$ resonance, $W=\sqrt{s}$  is the c.m. energy,  $M_c$ is the resonance mass, $\Gamma$ is the total width and $k$ is the center of mass (c.m.) momentum of the colliding particles. At the maximum of either of the $P_c$ resonances this expression gives numerically (at $k \approx 2.1\,GeV$)
\begin{equation}
\sigma_{max}(\gamma + p \to P_c \to J/\psi + p) \approx { 2 J +1 \over 4} \, Br(P_c \to \gamma + p) \, Br(P_c \to J/\psi +p) \, 1.1 \times 10^{-27} {\rm cm}^2~.
\label{bwm}
\end{equation}

\begin{figure}[h]
\begin{center}
\includegraphics[width=0.7\textwidth]{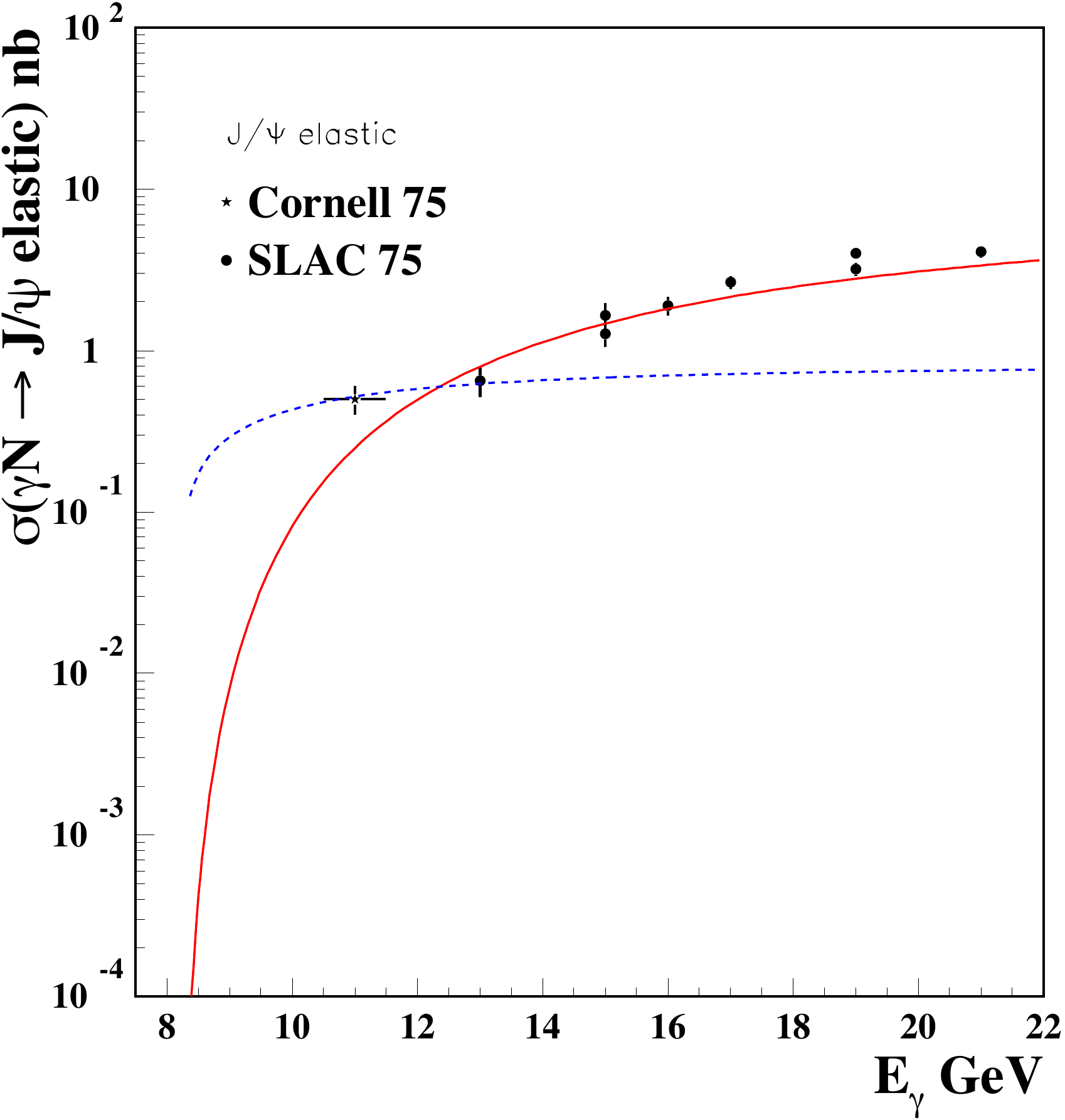}
\caption{
Variation of the
J/$\psi$ photoproduction cross section  %PH
near threshold. Solid line:  %PH I SEE NO SOLID LINE?
two gluon exchange. Dashed line: three gluon 
exchange~\cite{brodsky}.
}
\label{figbrod}
\end{center}
\end{figure} 

 The actual value of the cross section in Eq.(\ref{bwm}), naturally, depends on the product of the branching fractions, neither of which is presently known. However,  the branching fraction $Br(P_c \to \gamma + p)$ can be estimated in terms of $Br(P_c \to J/\psi + p)$ in a way that does not directly rely on a specific model of the internal dynamics of the pentaquarks, but which is somewhat sensitive to the structure of the amplitude of the decay  $P_c \to J/\psi +p$. 

It was shown in~\cite{vpk} that the bounds for the formation cross section at the resonance maximum
assuming the quantum numbers $J^P=(3/2)^-$ and $(5/2)^+$ 
respectively for $P_c(4380)$ and $P_c(4450)$ are:

\begin{eqnarray}
&& 1.5 \times 10^{-30}\, {\rm cm^2} \, < \, {\sigma_{max}[\gamma + p \to P_c(4380) \to J/\psi + p] \over Br^2[ P_c(4380) \to J/\psi+p]} \, < \, 47 \times 10^{-30} \, {\rm cm^2}~, \nonumber \\
&& 1.2 \times 10^{-29}\, {\rm cm^2} \, < \, {\sigma_{max}[\gamma + p \to P_c(4450) \to J/\psi + p] \over Br^2 [ P_c(4450) \to J/\psi+p]} \, < \, 36 \times 10^{-29} \, {\rm cm^2}~,
\label{modn}
\end{eqnarray}
where the lower bound corresponds to the presence of only the lower allowed partial wave, while the upper bound is found in the opposite situation where only the higher orbital wave is present.  Currently neither branching fraction $Br(P_c \to J/\psi +p)$ is known, nor the spin-parity quantum numbers for the observed pentaquarks are determined with certainty. Thus, it is not possible to estimate more definitely the value of the discussed formation cross section in a model independent way.

\begin{figure}[h]
\begin{center}
\includegraphics[width=1.1\textwidth]{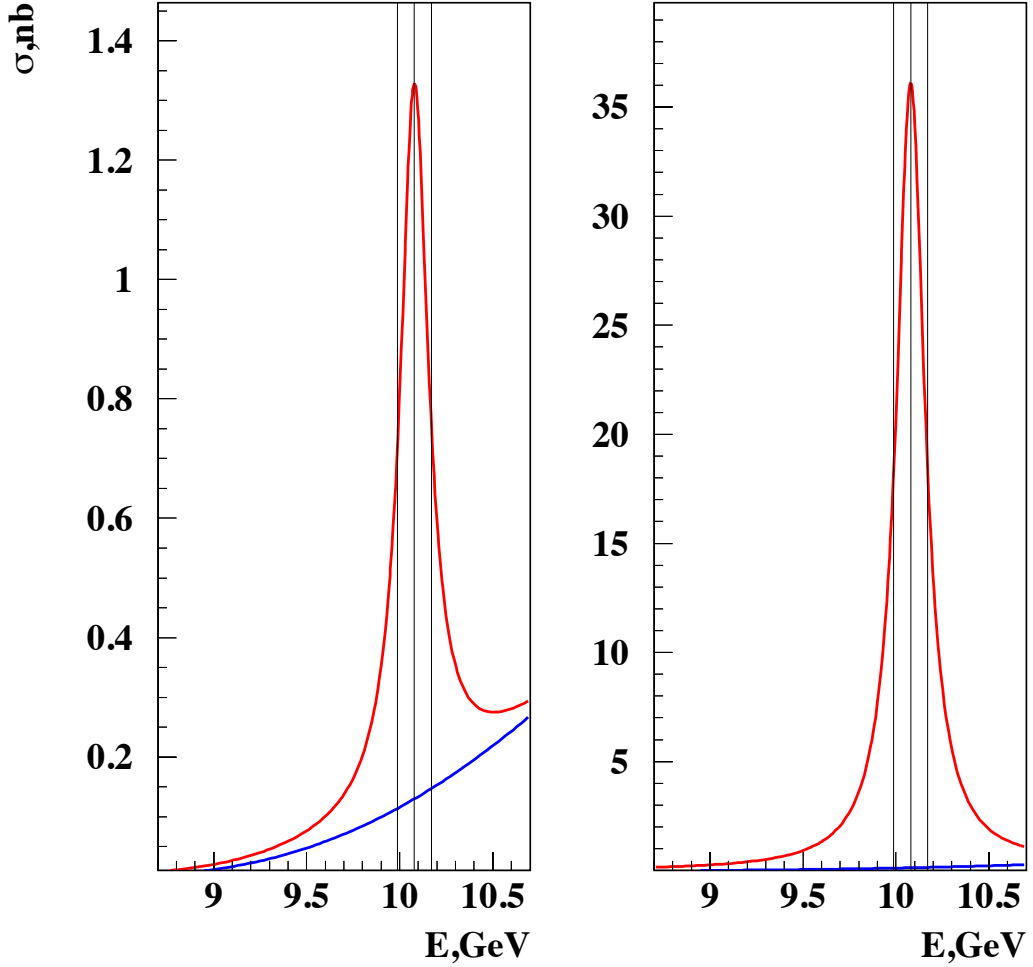}
\vskip -6.0cm
\caption{The $P_c(4450)$ resonance formation cross section in the reaction $\gamma p\to P_c \to J/\psi p$ as a function of the photon energy in the region of the CLAS12 acceptance. Two panels represent the theoretical uncertainty due to the unknown composition of the partial waves (see text for details). 
The vertical lines represent the resonant energy $E_0=10.1$~GeV and the boundaries  of the 
region $M_c\pm\Gamma/2$  (see Eq.~\ref{bwm}) in the laboratory system.
The two curves show the elastic background~\cite{brodsky}  and Breit-Wigner distribution.
The calculations were done assuming $Br(P_c\to J/\psi p)=1\%$. The cross section is proportional to
$Br^2(P_c\to J/\psi p)$. 
}    
\label{figbw}
\end{center}
\end{figure} 

 The $J/\psi$ photoproduction cross section is presented in Fig.~\ref{figbrod} as a function of the photon beam energy in the laboratory system~\cite{brodsky}. There are no published experimental data around $E_\gamma=10$ GeV. 
  We used the theoretical curve from ~\cite{brodsky} to estimate the elastic background under the pentaquark peaks. 
 It is shown as red line in Fig.~\ref{figbrod}.
The model-dependent extrapolation to the region of interest gives the cross section $\sim0.1$~nb.

 Fig.~\ref{figbw}  illustrates the  $P_c(4450)$ pentaquark formation cross section  as a function of the photon beam energy in the CLAS12 acceptance region for two cases. The left panel 
corresponds to the presence of only the lower allowed partial wave. The right panel shows the same cross section for  the upper bound where only the higher orbital wave is present (see Eq.~\ref{modn}).
 The vertical lines represent the resonant energy $E_0=10.1$~GeV and the boundaries of the 
region $M_c\pm\Gamma/2$  (see Eq.~\ref{bwm}) in the laboratory system that contains 50\% of the total production cross section.
The calculations were done with $Br(P_c\to J/\psi p)=1\%.$ The cross section is proportional to $Br^2(P_c\to J/\psi)$.  If the $J/\psi$ meson will be detected in the $e^+e^-$ or $\mu^+\mu^-$ decay mode we need to take into account the branching ration $Br(J/\psi \to e^+e^-)\sim Br(J/\psi \to \mu^+\mu^-)=6\%$.

\section{Estimate of the pentaquark yield}

The total number of the events in CLAS12 for the reaction $\gamma p\to P_c \to J/\psi p$, $J/\psi\to e^+e^-$ was estimated as
\begin{eqnarray}
N(P_c) = \int \sigma(W) \frac{dL_\gamma}{dW}\  \epsilon\ dW \cdot Br(J/\psi\to e^+e^-)=0.5\int \sigma(W) dW
\mbox{ events/nb/MeV/day}\label{modn}.
\end{eqnarray}
\noindent
where $\frac{dL_\gamma}{dW}=10^{30}$ events/cm$^2$/MeV/s=10$^{-3}$ events/nb/MeV/s is the CLAS12 photon luminosity 
per $W=1\ MeV$ bin at maximum electron-proton luminosity $L_{ep}=10^{35}$ events/cm$^2$/s
calculated at $E_\gamma\sim$10 GeV~\cite{stepan}, $\epsilon=0.1$ is the CLAS12 acceptance 
% at $E_\gamma\sim$10 GeV~\cite{stepan} 
 and $Br(J/\psi \to e^+e^-)=6\%$. 
 The experimental details  are described in the proposal \cite{E12-12-001}.
 For elastic $J/\psi$ photoproduction we have $\sigma=0.1$ nb. Integrated over 20 MeV bin yield is 1 event/day. 

The pentaquark yield for one day at this luminosity is presented in the Table~\ref{tab1} for
two states and two values of the predicted cross sections shown in Fig.~\ref{figbw},  assuming $Br(P_c \to J/\psi p)=1\%$. The theoretical bounds for the cross section are listed in Eq.~\ref{modn} for $P_c(4380)$ and $P_c(4450)$ states.
The elastic background is small even for the lower bounds of the cross section.

\begin{table}[h]
\centering
\caption{Estimated number of detected by CLAS12 events per day.}
\label{tab1}
\begin{tabular}{|c | c | } 
\hline
                      &  Number of events per day, minimum-maximum \\ \hline 
$P_c(4380)$ & 24\ \ \ -\ \ \ \ 750   \\ \hline 
$P_c(4450)$ &  35\ \ \ -\ \ \ 1100  \\ 
\hline 
\end{tabular}
\end{table}

\section{Conclusion}

Resolving between the models and clarifying the nature of the discovered hidden-charm pentaquark peaks, and possibly searching for similar peaks with other quantum numbers, requires further experimental studies. 
These states were observed in the  decay mode $J/\psi+p$. Thus, it is natural to expect that these pentaquarks can be produced in the photoproduction process $\gamma+p\to P_c\to J/\psi+p$ where these states will appear as s-channel resonances at photon energy around 10\,GeV. The energy and luminosity of the CLAS12 photon beam permit the
detailed studies of the production and decay properties of the pentaquark resonances. By this reason the pentaquark  search at Jefferson Laboratory looks extremely attractive.

The peak cross section for $\gamma + p \to P_c \to J/\psi + p$
 is likely to reach values that
may allow fairly detailed studies of the pentaquarks and a search for other similar states. In particular, it may be realistic to study the decays of the $P_c$ states into $J/\psi p \pi$ and $J/\psi p \pi \pi$. 
Such decays should be prominent, if the $P_c$ states are dominantly a baryo-charmonium, (i.e. a hadro-quarkonium) types~\cite{mv07,dv} composed of $J/\psi$ and excited nucleon states similar to the known resonances $N(1440)$ and $N(1520)$. 
The CLAS12 detector has enough acceptance to detect these decay modes.
These  patterns of the decays of the $P_c$ resonances would disfavor the molecular models~\cite{cllz,cclsz,rno,he}, where one would expect the natural decay channels into a charmed hyperon and a meson, or from the $\chi_{c1} p$ complex model~\cite{mo}, where the expected dominant decay is $P_c(4450) \to \chi_{c1}+p$. Naturally, any observation of the $P_c$ peaks in the $\gamma p$ cross section would 
confirm the resonance nature of
the peaks and rule out the interpretation~\cite{gmwy,lwz,mikhasenko,amnss} as
 `accidental' singularities in the $\Lambda_b$ decays.

 The work of V.K. is supported by the U.S. Department of Energy.  The Jefferson Science Associates (JSA) operates the Thomas Jefferson National Accelerator Facility for the United States Department of Energy under contract DE-AC05-06OR23177. 
 The work of M.B.V.  is supported in part by U.S. Department of Energy
Grant No. DE-SC0011842, and was performed in part at the Aspen Center
for Physics, which is supported by National Science Foundation grant 
PHY-1066293.

%\bibliography{3dstructure}

\end{document}